\documentclass[3p,times,twocolumn]{elsarticle}

\usepackage{ecrc-mb}
\usepackage{widetext}

\volume{00}

\firstpage{1}

\journalname{Nuclear Physics B Proceedings Supplement}

\runauth{M. Beneke}
\runtitle{Unstable-particle effective field theory}
\runprep{TUM-HEP 975/15, SFB/CPP-14-118, 
27 December 2014}

\jid{nuphbp}

\jnltitlelogo{Nuclear Physics B Proceedings Supplement}

\usepackage{amssymb}

\usepackage[figuresright]{rotating}

\begin{document}

\begin{frontmatter}



\dochead{}


\title{Unstable-particle effective field theory}


\author{M. Beneke}

\address{Physik Department T31, 
James-Franck-Stra\ss e~1, 
Technische Universit\"at M\"unchen,
D--85748 Garching, Germany}

\begin{abstract}
\noindent
Unstable particles are notorious in perturbative quantum field theory 
for producing singular propagators in scattering amplitudes that 
require regularization by the finite width. In this review I discuss 
the construction of an effective field theory for unstable particles, 
based on the hierarchy of scales between the mass, $M$, and the width,
$\Gamma$, of the unstable particle that allows resonant 
processes to be systematically expanded in powers of the coupling $\alpha$
and $\Gamma/M$, thereby providing gauge-invariant approximations 
at every order. I illustrate the method with the 
next-to-leading order line-shape of a scalar resonance in an abelian
gauge-Yukawa model, and results on NLO and dominant NNLO corrections 
to (resonant and non-resonant) pair production of $W$-bosons and 
top quarks.
\end{abstract}

\begin{keyword}
Unstable particles, effective field theory, perturbative quantum 
field theory, line-shape, top quark, $W$-boson 


\end{keyword}

\end{frontmatter}


\section{Introduction}
\label{sec:intro}

The consistency of the Standard Model (SM) of particle physics is 
tested at high-energy colliders primarily through the production and 
subsequent decay of unstable particles. New particles, if discovered, 
are most likely also short-lived. At the current high-energy frontier 
all known fundamental interactions are perturbatively weak, allowing 
for very precise theoretical computations in principle. Nevertheless, 
the application of perturbation theory to processes with unstable 
particles is not always straightforward.

The very notion of an unstable particle requires clarification.
In quantum field theory the fundamental entities are the fields from
which the Lagrangian is constructed, but the excitations of the
fundamental fields may not correspond to the asymptotic particle
states assumed in scattering theory, if they are strongly interacting 
as is the case for the quarks and gluons of QCD, or unstable with 
respect to decay into lighter particles. Relevant cases include the 
electroweak gauge bosons and the top quark, which although all very 
short-lived, have width over mass ratios of a few percent, larger than 
the accuracy of precision calculations. Principal questions related to field
theories with unstable ``particles'' such as their unitarity on the
Hilbert space built upon the one-particle states of only stable particles 
have been answered many years ago \cite{Veltman:1963th}. The construction 
of the unitary $S$-matrix is based on certain properties of the exact 
two-point function of the unstable-particle field. Although a diagrammatic 
interpretation is assumed, there is no explicit reference to a 
perturbation expansion in the coupling that renders the particle unstable.

Since exact two-point functions are not at hand, this raises the 
question of consistent, successive approximations. Ordinary perturbation 
theory in the Lagrangian coupling $g$ does not work, since the 
lowest-order propagator of the unstable particle leads to singularities 
in scattering amplitudes. A well-known remedy of the singularity is  the 
resummation of self-energy corrections 
to the propagator, which results in the substitution
\begin{equation}
\frac{1}{p^2-M^2}\,\to\,\frac{1}{p^2-M^2-\Pi(p^2)}.
\end{equation} 
The self-energy has an imaginary part of order $M^2 g^2 \sim 
M\Gamma$, where $\Gamma$ is the on-shell decay width of the 
resonance, rendering the propagator large but finite. ``Dyson 
resummation'' sums a subset of singular terms of order 
$(g^2 M^2/[p^2-M^2])^n\sim 1$ (near resonance where $p^2\sim M\Gamma$) 
to all orders in the expansion in $g^2$. This procedure 
leaves open the question of how to identify 
all terms (and only these) required to achieve a specified 
accuracy in $g^2$ and $\Gamma/M$. Failure to address this 
question may lead to a lack of gauge invariance and unitarity of 
the resummed amplitude, since these properties are guaranteed only 
order-by-order in perturbation theory, and for the exact 
amplitude. 

Despite the fact that unstable particle fields have no corresponding 
asymptotic particle states and hence their propagators should never 
be cut, this point is often ignored in practice 
and the particle is treated in Feynman diagram and cross section 
calculations as if it were stable (``narrow-width approximation''). 
This can be justified when the width is very small, since 
\begin{equation}
\frac{M \Gamma}{(p^2-M^2)^2+M^2\Gamma^2} \stackrel{\Gamma\to 0}{\to}
\pi\delta(p^2-M^2).
\label{eq:deltafnapprox}
\end{equation} 
The limit holds in the distribution sense and is therefore valid 
only, if the phase-space of the unstable particle is integrated sufficiently 
inclusively, such that the integration contour in the variable $p^2$ 
can be deformed far away from $M^2$. This is not always the case. 
An obvious example is the line-shape, but also distributions may 
trap the contour near $M^2$. A more accurate treatment than 
(\ref{eq:deltafnapprox}) is also required, when the desired precision 
exceeds the leading-order approximation in $\Gamma/M$.

Somewhat surprisingly, systematic
computational schemes to obtain approximations to scattering processes
involving unstable particles in weak coupling expansions are relatively 
recent. The two, which are general, are the 
{\em unstable-particle effective theory} and the 
{\em complex mass scheme}.

\subsection{Unstable-particle effective theory}

The singularity of the unstable particle propagator indicates 
sensitivity to a time scale larger than the Compton wave-length $1/M$ 
of the particle, evidently its lifetime $1/\Gamma$. The presence of 
two different scales $\Gamma\ll M$ lies in the very nature of the 
problem, since a resonance with $\Gamma \sim M$ would not be identified 
as such. The main idea of unstable-particle effective field theory  
\cite{Beneke:2003xh,Beneke:2004km} is to exploit this hierarchy of 
scales in order to systematically
organize the calculations in a series in the coupling $g$, and
$\Gamma/M$. The short-distance scale $M$ is integrated out by 
performing standard perturbative computations and the full theory 
is matched to an effective Lagrangian that reproduces the physics at the 
scale $\Gamma$. The effective theory contains a field $\phi_v$, 
which describes a resonance with momentum $p=M v+k$, where only 
$k\sim \Gamma$ is fluctuating. The resonant field can interact with other soft 
fields with momenta of order $\Gamma$, but off-shell effects at the 
scale $M$ are part of the matching coefficients. The power-counting of 
fields and interactions in the effective theory leads to a systematic 
construction of the expansion in $\Gamma/M$. 

The expansion of amplitudes in matching calculations is performed 
around the gauge-invariant location 
\begin{equation}
M_\star^2 = M^2 - i M\Gamma
\label{eq:complexpole}
\end{equation}
of the pole in the complex $p^2$ plane corresponding to the resonance, 
where $M$ is identified with the pole mass, and $\Gamma$ with the 
on-shell width. The expansion is similar to the one performed in the 
``pole'' \cite{Stuart:1991xk,Aeppli:1993rs} or ``double-pole'' 
(in pair production of resonances) \cite{Beenakker:1998gr,Denner:1999kn} 
approximation. In a certain sense, unstable-particle effective field 
theory represents the field-theoretic formulation of the diagrammatic 
pole approximation, and generalizes it to all orders in perturbation 
theory and beyond the leading power in the $\Gamma/M$ expansion. A first
step in this direction had already been presented 
in \cite{Chapovsky:2001zt}. 

The effective theory approach is minimal as it identifies precisely 
the terms required to achieve a specified accuracy in $g^2$, and
$\Gamma/M$, and does not include more. This makes the calculations 
particularly simple. Furthermore, the operator interpretation 
allows for the summation of large logarithms of $\Gamma/M$ through 
renormalization group equations and anomalous dimensions. There is 
a draw-back: the details of the effective theory depend on the 
inclusiveness of the observable and is not valid locally over the entire 
phase-space, where some portions may involve further soft scales 
of order $\Gamma$ in addition to $(p^2-M^2)/M$. Even the prediction 
of the resonance line-shape requires matching of the resonant (peak) 
region calculation within the effective theory to the off-resonance 
region computed with standard perturbation theory.

\subsection{The complex-mass scheme}

The complex-mass scheme is an extension of the standard on-shell 
renormalization scheme to unstable particles. It defines the complex mass 
and field renormalization constant from the location and 
residue of the pole (\ref{eq:complexpole}) of the unstable-particle 
propagator. The bare mass $M_0$ is split into a renormalized mass and 
counterterm through
\begin{equation}
M_0^2 = M_\star^2 +\delta M_\star^2,
\label{eq:cms}
\end{equation}
and $\delta M_\star^2$ is part of the interaction Lagrangian and treated 
as a perturbation. The 
unstable-particle propagator $i/(p^2-M_\star^2)$ is never infinite for 
physical, real momenta. The complex-mass scheme was discussed already in 
\cite{Stuart:1990vk}, but it was used for the first time in a full 
one-loop calculation only in 2005 \cite{Denner:2005fg} for the process 
$e^+ e^- \to 4\,\mbox{fermions}\,(+\gamma)$ at high energies, which 
receives important contributions from the unstable $W^+ W^-$ intermediate 
state.

Although the standard rules of perturbation theory and an expansion in 
the number of loops apply to the complex-mass scheme, a re-ordering and 
resummation of the $g^2$ expansion is implicit, since the propagator is 
of order $1/(M^2 g^2)$ in the resonance region. The assumption is that 
the complex mass in the propagator captures all terms that need to 
be resummed which is indeed the case (see also next section). Since the 
scheme is only a reparameterization of the bare theory, which is not 
modified, it is obvious that no double counting occurs. Likewise, gauge 
invariance is assured, since the split (\ref{eq:cms}) is gauge-invariant 
and the algebraic identities that guarantee gauge invariance are valid 
in the presence of complex parameters. Unitarity might be a concern, 
since the unitarity equation involves complex conjugation. However, since 
the bare theory is unitary, so must be the reparameterized one. What needs 
to be shown is that the theory with the complex-mass prescription 
is perturbatively unitary in the sense that unitarity violation in any 
given order in the loop expansion are of higher order in the expansion 
parameters (counting $\Gamma/M\sim g^2$). This point was demonstrated 
explicitly at one-loop for fermion-fermion scattering through a 
vector-boson resonance~\cite{Bauer:2012gn}, and in general 
in \cite{Denner:2014zga}.

The complex-mass scheme is conceptually straightforward. It does not 
require separate treatments of the resonance and off-resonance regions, 
and can easily be applied to kinematic distributions. Compared 
to the effective field theory method the scheme does not make use 
(explicitly) of 
expansions in $\Gamma/M$ and hence does not simplify the problem 
as much as possible in principle. The difficulty of the calculation 
is equivalent to the corresponding standard loop calculation with the 
additional complication of loop integrals with complex masses. This 
is not a practical problem at the one-loop order, making the 
complex-mass scheme the method of choice for automated next-to-leading 
order calculations. On the other hand, calculations beyond this 
order would presently be difficult and the resummation of 
logarithms $\ln M/\Gamma$ cannot be performed.

In the following I do not discuss the complex-mass scheme further, but 
focus on unstable-particle effective theory. I use the line-shape of 
a resonance to illustrate the framework and the discuss results on 
pair production of W-bosons and top quarks near threshold which (I believe) 
benefit particularly from this method.

\section{Line-shape of an unstable particle}

In this section, which follows \cite{Beneke:2003xh,Beneke:2004km}, 
we consider a toy model that involves a massive 
scalar field, $\phi$, and two fermion fields. The scalar as well 
as one of the fermion
fields, $\psi$, (the ``electron'') are charged under an abelian gauge
symmetry, whereas the other fermion, $\chi$, (the ``neutrino'') is
neutral. The model allows for the scalar to decay into an
electron-neutrino pair through a Yukawa interaction. The model describes 
the essential features of the $Z$-boson line-shape in the SM 
\cite{Bardin:1989qr}. Its 
Lagrangian is
\begin{eqnarray}
\label{model}
{\cal L} &=& (D_\mu\phi)^\dagger D^\mu\phi - \hat M^2 \phi^\dagger\phi +
 \bar\psi i \!\not\!\!D\psi + \bar\chi i\!\!\not\!\partial\chi
\nonumber \\
&& - \, \frac{1}{4} F^{\mu\nu}F_{\mu\nu}-\frac{1}{2\xi} \,
(\partial_\mu A^\mu)^2
 \nonumber\\
 && + \, y\phi\bar\psi\chi + y^* \phi^\dagger \bar\chi\psi
-\frac{\lambda}{4}(\phi^\dagger \phi)^2+ {\cal L}_{\rm ct}\, ,
\end{eqnarray}
where $\hat{M}$ denotes the renormalized mass, not necessarily the pole  
mass $M$ defined by (\ref{eq:cms}), ${\cal L}_{\rm ct}$ the
counterterm Lagrangian, and $D_\mu=\partial_\mu-i g A_\mu$. 
We define $\alpha_g\equiv
g^2/(4\pi)$, $\alpha_y\equiv (y y^*)/(4 \pi)$ (at the scale $\mu$) and assume
$\alpha_g \sim \alpha_y \sim \alpha$, and $\alpha_\lambda\equiv\lambda/(4\pi)
\sim \alpha^2/(4\pi)$.

The line-shape is the totally inclusive cross section for the
process 
\begin{equation}
\bar\nu(q) + e^-(p)\to X
\label{process}
\end{equation}
as a function of $s\equiv (p+q)^2$, which can be calculated from the 
imaginary part of the forward scattering amplitude 
${\cal T}(s)$.\footnote{The total cross section of process (\ref{process}) 
is not infrared finite for massless electrons due to an initial-state
collinear singularity, which has to be absorbed into the electron
distribution function. In what follows it is understood that this
singularity is subtracted minimally.}  In particular, we are
interested in the region $s\approx M^2$, or more precisely $s-M^2 \sim
M\Gamma\sim\alpha M^2 \ll M^2$, where we expect an enhancement of the cross
section due to the resonant production of the scalar. Defining the 
dimensionless variable
\begin{equation}
\label{eq:deltadef}
\delta\equiv \frac{s-\hat M^2}{\hat M^2} \sim \frac{\Gamma}{M},
\end{equation}
the cross section far away from the resonance can be expanded in $g^2$ 
in the usual manner according to
\begin{equation}
\sigma =g^4f_1(\delta)+g^6 f_2(\delta)+\ldots.
\end{equation}
At every order, the coefficient $f_n(\delta)$ is a function of the 
variable $\delta \sim 1$. On the other hand, near resonance we may 
exploit $\delta \ll 1$ to expand the amplitude in $\delta$. At the same 
time, as $g^2/\delta\sim 1$ since $\Gamma\sim M g^2$, some terms 
must be summed to all orders. A systematic approximation to the line-shape 
in the resonance region therefore takes the form
\begin{eqnarray}
\sigma &\sim& \sum_n \left(\frac{g^2}{\delta}\right)^{\!n}\!\times 
\{1 \,{\rm (LO)}; g^2,\delta\, {\rm (NLO)}, \ldots\} 
\nonumber\\
&=& 
h_1(g^2/\delta)+ g^2 h_2(g^2/\delta)+\ldots
\label{eq:reorganizedexp}
\end{eqnarray}
with non-trivial functions $h_n(g^2/\delta)$ at every order in the 
reorganized expansion. The effective theory identifies the relevant terms 
and constructs the expansion (\ref{eq:reorganizedexp}).

\subsection{Relevant modes and reduced scattering diagrams}

The effective theory is based on the hierarchy of scales $\Gamma\ll M$. 
In a first step we integrate out hard momenta $k\sim M$. The effective 
theory will then not contain any longer dynamical hard modes since 
their effect is included in the coefficients of the operators. The hard 
effects are associated with what is usually called factorizable corrections, 
whereas the effects of the dynamical modes correspond to the 
non-factorizable corrections \cite{Chapovsky:2001zt}. On the level of 
Feynman diagrams, the hard contribution can be identified directly 
using  the method of regions to separate loop integrals into various
contributions~\cite{Beneke:1997zp}. The hard part is obtained by
expanding the full-theory integrand in $\delta$. 

The modes to be described by the effective Lagrangian correspond to 
kinematically allowed scattering processes with virtualities much 
smaller than $M^2$. Particles with masses above $M\Gamma$ 
are no longer present, except for the unstable particle, which by 
construction is close to mass-shell. To account for this, we write 
the momentum of the scalar particle as $P=\hat{M}v + k$, where the velocity
vector $v$ satisfies $v^2=1$ and the residual momentum $k$ scales as
$M\delta \sim \Gamma$.  In analogy to heavy-quark effective theory (HQET) 
we remove the rapid
spatial variation $e^{-i\hat{M} v\cdot x}$ from the $\phi$ field and
define $\phi_v(x) \equiv e^{i\hat{M} v\cdot x}\, {\cal P}_+ \phi(x)$,
where ${\cal P}_+$ projects onto the positive frequency part to ensure
that $\phi_v$ is a pure destruction field. A field with momentum 
fluctuations $k\sim \Gamma$ is called a ``soft'' field.  Thus, for the 
soft scalar field $\phi_v$ we have $P^2-\hat{M}^2 \sim M^2\delta$. This
remains true if the scalar particle interacts with a soft gauge boson
with momentum $M\delta$, so the effective Lagrangian should contain 
soft (s) fields for every massless field of the full theory. 

The unstable particle is produced in the scattering of on-shell particles 
with large energy of order $M$. These can remain near mass-shell by radiating 
further energetic particles in their direction of flight. The effective 
Lagrangian must therefore also contain hard-collinear (c1) modes with 
momentum scaling 
\begin{equation} 
n_+ p \sim M, \quad p_\perp \sim M\delta^{1/2}, 
\quad n_- p \sim M\delta 
\end{equation}
for all massless fields of the original Lagrangian. Here 
$n_\pm$ are two light-like vectors with $n_+\cdot n_-=2$, 
$n_-$ is the direction of the electron four-momentum, and $p_\perp$ 
is transverse to $n_-$ and $n_+$.\footnote{In the general case several 
types of collinear modes are required, one for 
each direction defined by energetic particles in the initial and 
final state. For the inclusive line-shape we calculate 
the forward-scattering amplitude, so no direction is distinguished in the 
final state. We then need two sets of collinear 
modes, one for the direction of the incoming electron, labelled 
by ``c1'' (or often simply ``c''), 
the other for the direction of the incoming neutrino (labelled ``c2''). 
Since the neutrino is electrically neutral, the collinear 
fields $\psi_{c2}$, $A_{c2}$ and $\chi_{c1}$ appear 
only in highly suppressed terms, so we can ignore them here.}

\begin{figure}
\vskip0.2cm
\begin{center}

\hskip-0.3cm
\includegraphics[width=7.8cm]{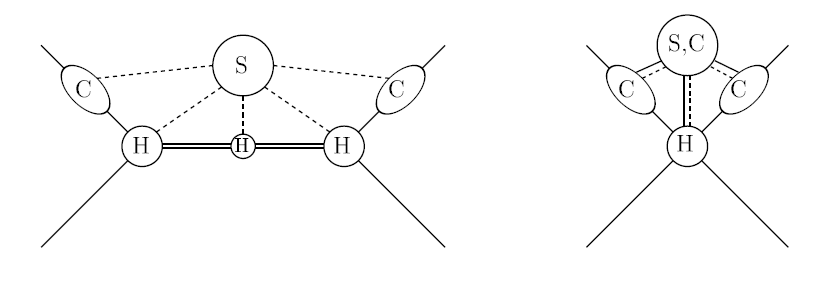}
\end{center}
\vspace*{-0.3cm}
\caption{\label{fig:skeleton}
Reduced diagram topologies in $2\to2$ scattering 
near resonance. Left: resonant scattering. Right: non-resonant  
scattering. }
\end{figure}

The space-time picture of the kinematically allowed processes is 
very simple and the corresponding reduced diagram topologies are 
shown in Figure~\ref{fig:skeleton} for the forward-scattering 
amplitude. The left diagram describes the production of the 
resonance through a hard process, represented in the effective theory 
by some local operator ${\cal O}_p^{(k)}$, and its subsequent 
propagation over distances of order $1/\Gamma$. The resonance (double 
line) can interact with soft fluctuations. The initial-state electron leg 
can be dressed with collinear corrections. However, collinear 
modes cannot be exchanged across the double line, since this would 
not leave enough energy to produce the scalar near resonance. 
The process just described is represented in the effective theory by 
the first line of the master formula 
\begin{eqnarray}
i \,{\cal T} &=& \sum_{k,l} \int d^4 x \,
\langle \nu e |T(i {\cal O}_p^{(k)}(0)i{\cal O}_p^{(l)}(x))|\nu e
\rangle
\nonumber\\ 
&& +\, \sum_{k} \,\langle \nu e|i {\cal O}_{\rm nr}^{(k)}(0)|\nu e\rangle.
\label{eq:master}
\end{eqnarray}
for the forward-scattering amplitude.

The scattering may also occur without the production of the scalar near its 
mass-shell  (right diagram in Figure~~\ref{fig:skeleton}). In the present 
toy theory this still requires an intermediate
scalar line, since the neutrino has only Yukawa interactions. 
The scalar may be off-shell, because the electron has radiated 
an energetic (hard or collinear) photon before it hits the neutrino. 
In this case the invariant mass of the colliding electron-neutrino system 
is of order $M^2$ but not near $M^2$, producing a non-resonant
scalar. In the effective theory this process is represented by 
a local four-fermion operator ${\cal O}_{\rm nr}^{(k)}$, without 
$\phi_v$ fields. In general, non-resonant scattering includes all 
``background processes'', which produce one of the final states under 
consideration. This topology 
does not involve a resonant heavy scalar, and both
soft and collinear fields can be exchanged across the diagram. 
The matrix elements in (\ref{eq:master}) are understood 
to be evaluated with the effective Lagrangian.

\subsection{Construction of the effective Lagrangian}

We divide the effective Lagrangian into three parts.  Roughly speaking, 
the first, ${\cal L}_{\rm HSET}$, describes the heavy scalar field near
mass-shell and its interaction with the gauge field. The second part,
${\cal L}_{\pm}$, describes energetic fermions and
their interactions with the gauge field. Finally, the third part,
${\cal L}_{\rm int}$ contains the local operators ${\cal O}_p^{(k)}$ 
and ${\cal O}_{\rm nr}^{(k)}$ responsible for the production of the 
resonance and off-shell processes. In the following, we write down 
all terms needed for a next-to-leading order (NLO) 
calculation of the line-shape.

The soft Lagrangian ${\cal L}_{\rm HSET}$ is an extension of the 
HQET Lagrangian \cite{Eichten:1989zv} to a (here scalar) particle whose 
mass-shell is defined by the complex pole location (\ref{eq:cms}). 
The residual mass term which is usually set to zero in HQET by choosing 
$M$ to be the pole mass of the heavy quark, is now necessarily 
non-vanishing and complex. The relevant terms are 
\begin{eqnarray}
{\cal L}_{\rm HSET} &=&  2 \hat{M}  \phi_v^\dagger\,
        \left( i v \cdot D_s - \frac{\Delta^{(1)}}{2} \right) \phi_v
\nonumber \\
&& \hspace*{-1cm} +\,2 \hat{M}  \phi_v^\dagger\,
        \left( \frac{(i D_{s,\top})^2}{2\hat{M}} +
               \frac{[\Delta^{(1)}]^2}{8\hat{M}} -
               \frac{\Delta^{(2)}}{2} \right) \phi_v
\nonumber  \\
&& \hspace*{-1cm}-\,\frac{1}{4} F_{s\mu\nu} F_s^{\mu\nu}
+\bar\psi_s i\!\not\!\!D_s \psi_s
+\bar\chi_s i\!\not\!\partial \chi_s, 
\label{eq:heavyNLO}
\end{eqnarray}
where $\psi_s$ ($\chi_s$) denotes the soft electron (neutrino) field 
and the covariant derivative $D_s\equiv \partial -ig A_s$ includes the 
soft photon field. Furthermore, $a^\mu_\top\equiv a^\mu-(v\cdot a)\, v^\mu$ 
for any vector. The only non-trivial short-distance matching 
coefficients in this 
expression are the quantities $\Delta^{(i)}$ to be defined below.

The bilinear terms in the soft scalar field $\phi_v$ are determined 
by the requirement that ${\cal L}_{\rm HSET}$ reproduces the two-point 
function of the scalar in the full theory close to resonance. Denoting the
complex pole of the propagator by $M_\star^2$ and the residue at the
pole by $R_\phi$, the propagator near resonance 
can be written as
\begin{equation}
\frac{i\, R_\phi}{P^2 - M_\star^2} =
\frac{i\, R_\phi}{2 \hat{M} v \cdot k + k^2 -(M_\star^2 -\hat{M}^2)}.
\label{eq:propagator}
\end{equation}
We now define the matching coefficient 
$\Delta \equiv (M_\star^2 -\hat{M}^2)/\hat{M}$. There are two
solutions to $P^2=M_\star^2$, one of which is irrelevant since it scales as
$v\cdot k\sim\hat{M}$. For the other we find
\begin{eqnarray}
v\cdot k &=& -\hat{M}+\sqrt{\hat{M}^2+\hat{M}\Delta-k_\top^2} \nonumber \\
&=& \frac{\Delta}{2}-\frac{\Delta^2+4 k_\top^2}{8\hat{M}}
 + {\cal O}(\delta^3) ,
\label{eq:propexp}
\end{eqnarray}
where we expanded in $\delta$ in the second line, using $\Delta \sim 
k_\top \sim M\delta$. Expanding $\Delta = \sum_{i=1}  \Delta^{(i)}$ into 
terms of order $g^{2 i}$, we deduce the bilinear terms in (\ref{eq:heavyNLO}) 
from the dispersion relation (\ref{eq:propexp}). Gauge invariance of 
the effective Lagrangian implies that the leading soft-photon interactions 
can be obtained from the bilinear terms by replacing
$\partial\to D_s$. The gauge invariance of the matching coefficient 
follows from the invariance of the unstable-particle pole $M_\star$.

In the underlying theory the full renormalized propagator of the 
unstable particle is given by 
$i (s-\hat{M}^2-\Pi(s))^{-1}$, where $-i\,\Pi(s)$ corresponds to the 
amputated 1PI graphs including counterterms. Comparing this to 
(\ref{eq:propagator}) and expanding $\Pi(s)$ around $\hat M^2$ 
and in the number of loops in the form $\Pi(s) = \hat{M}^2\, 
\sum_{k,l} \delta^l \, \Pi^{(k,l)}$, where it is understood that 
$\Pi^{(k,l)}\sim g^{2k}$, we obtain
\begin{equation}
  \Delta = 
   \hat{M}\, \Pi^{(1,0)} +
   \hat{M}  \left(\Pi^{(2,0)}+\Pi^{(1,1)}\Pi^{(1,0)} \right) + \ldots.
  \label{eq:Deltaexp}
\end{equation}
$\Pi^{(1,0)}$ and $\Pi^{(2,0)}+\Pi^{(1,1)}\Pi^{(1,0)}$ (but not 
$\Pi^{(2,0)}$ and $\Pi^{(1,1)}$ separately) are infrared-finite, which 
justifies the interpretation of $\Delta$ as a short-distance coefficient. 
Explicit results for $\Delta^{(1)}$ and $\Delta^{(2)}$ in the
$\overline{\rm MS}$ and pole renormalization scheme can be found in
\cite{Beneke:2004km}. 
Here we only note that in the pole scheme ($\hat{M} \equiv M$), we have 
$\Delta = -i\Gamma$, in which case the residual ``mass'' is purely 
imaginary and coincides with the on-shell width. 

Each term in ${\cal L}_{\rm HSET}$ can be assigned a scaling power in
$\delta$. Since $D_s \sim k \sim \Gamma\sim M\delta$ and 
$\Delta^{(1)}\sim M g^2\sim M\delta$, both terms
in the first line of (\ref{eq:heavyNLO}) are of equal size and 
leading terms. The unstable-particle propagator is therefore 
\begin{equation}
\frac{i}{2 \hat{M} (v\cdot k-\Delta^{(1)}/2)},
\label{eq:unstableprop}
\end{equation}
which corresponds to a fixed-width prescription. The linearity of the  
propagator in the (residual) momentum makes calculations in 
the effective theory particularly simple. The fact that only $\Delta^{(1)}$ 
appears in the leading-order Lagrangian proves that only the two-point 
function in the original theory needs to be resummed by including the 
one-loop self-energy into the unperturbed Lagrangian. 
No higher-point functions require resummation, which 
is intuitively obvious, since the origin of the long-distance scale 
is associated with a single-particle effect, the life-time of the 
resonance.

In momentum space the propagator (\ref{eq:unstableprop}) of the $\phi_v$ 
field scales as $1/\delta$. Hence, because $\int d^4 k$ counts as
$\delta^4$, the soft scalar field $\phi_v(x)$ scales as
$\delta^{3/2}$.  It follows that the terms in the second line of  
(\ref{eq:heavyNLO}) scale as $\delta^5$. Being suppressed by one
power in $\delta$ or $g^2$ relative to the first line, they must be 
included only in a calculation of the line-shape with NLO precision.  
Finally, 
since $A^\mu_s$ scales as $\delta$ and the soft fermion fields scale 
as $\delta^{3/2}$, the terms in the last line of (\ref{eq:heavyNLO})
scale as $\delta^4$ and represent leading interactions among the soft, 
massless modes. By adding further terms the Lagrangian can be improved 
to any accuracy desired.

Next, we turn to the construction of the effective Lagrangian, ${\cal
L}_\pm$, associated with the energetic fermions. The interactions of 
collinear modes with themselves and with soft modes are described within 
soft-collinear effective theory (SCET) 
\cite{Bauer:2000yr,Bauer:2001yt,Beneke:2002ph,Beneke:2002ni}. 
The coupling of collinear modes to the scalar field $\phi_v$, and 
among collinear fields with different directions produces off-shell 
fluctuations, which are not part of the effective Lagrangian. 
The momenta associated with generic collinear fields $\psi_{c1}$ and 
$\bar{\chi}_{c2}$ do not add up to a momentum of the form $P=M v+k$. 
This kinematic constraint is implemented by adding the 
production and non-resonant operators, ${\cal O}_{\rm p}^{(k)}$ and 
${\cal O}_{\rm nr}^{(k)}$, respectively, as external ``sources'' 
for the specific process. The line-shape is then given by 
the correlation function (\ref{eq:master}). 

Alternatively, the dynamical hard-collinear modes can be integrated 
out in a second matching step, in which the collinear functions 
(labelled ``C'' in Figure~\ref{fig:skeleton}) appear as matching 
coefficients of (non-local) operators. The new effective Lagrangian 
contains an ``external-collinear'' electron mode with momentum 
$\hat{M} n_-/2+k$, which describes the remaining soft fluctuations 
$k\sim\delta$ around the fixed large component. Similar to the resonance 
field, we extract the large component and define $\psi_{n_-}(x) 
\equiv e^{i\hat{M}/2\,(n_- x)}\, {\cal P}_+\,\psi_{c1}(x)$, where 
${\cal P}_+$ projects on the positive frequency part of $\psi_{c1}$. 
Adding the corresponding 
field with $n_-$ and $n_+$ exchanged for the neutrino, the 
soft interactions of the external-collinear field are given by  
\begin{equation}
{\cal L}_{\pm} =
\bar{\psi}_{n_-} \!i n_- D_s
\frac{\!\not\!n_+}{2} \, \psi_{n_-} +
\bar{\chi}_{n_+} \! i n_+ \partial\,
\frac{\!\not\!n_-}{2} \, \chi_{n_+} .
\label{eq:Lpm}
\end{equation}
at leading power.

With the external-collinear modes we can implement the production and
non-resonant sources as interaction terms in ${\cal L}_{\rm int}$. 
At NLO the relevant terms read
\begin{eqnarray}
{\cal L}_{\rm int} &=& C\, y\, \phi_v \bar{\psi}_{n_-}\chi_{n_+}
+ C\, y^* \phi_v^\dagger \bar{\chi}_{n_+}\psi_{n_-}
\nonumber \\
&+& D\, \frac{y y^*}{\hat{M}^2}
  \left(\bar{\psi}_{n_-} \chi_{n_+}\right)  \!
  \left(\bar{\chi}_{n_+} \psi_{n_-}\right) ,
\label{eq:Lint}
\end{eqnarray}
where $C=1+{\cal O}(\alpha)$ and $D$ are the matching
coefficients. The two lines correspond to the two reduced 
diagram topologies in Figure~\ref{fig:skeleton}. We note that 
the effective Lagrangian is not manifestly hermitian, since it 
describes the decay of the scalar. Nevertheless, it generates a 
unitary time evolution, since it reproduces by construction the unitary 
underlying theory to the specified order in the expansion in 
$\delta$.

The external fields scale as $\delta^{3/2}$. Thus, an
insertion of a $\phi\psi\chi$ operator results in $\int d^4x\, \phi_v
\bar{\psi}_{n_-}\chi_{n_+} \sim \delta^{1/2}$. The forward-scattering
amplitude requires two insertions of this operator. Accounting for the 
scaling of the external state $\langle \bar{\nu} e^-|
\sim \delta^{-1}$, we find  ${\cal T}^{(0)} \sim g^2/\delta$ for 
the amplitude at leading order, which is the expected result.
The four-fermion operator is suppressed in $\delta$ and results in a
contribution of order $g^2$ to ${\cal T}$. Thus, to compute the NLO 
correction ${\cal
T}^{(1)}$ we need $C^{(1)}$, the ${\cal O}(g^2)$ contribution to
the matching coefficient $C$, while $D$ is only needed at tree level. 
The matching coefficients are obtained from the hard contributions 
to the corresponding on-shell three- and four-point functions in 
the full theory. I refer to \cite{Beneke:2004km} for the precise matching
equation as well as the explicit results.

\subsection{Example diagram}

\begin{figure}
\vskip0.2cm
\begin{center}
\includegraphics[width=2.3cm]{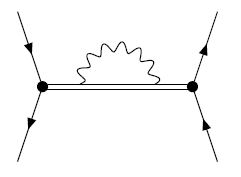}
\end{center}
\vspace*{-0.3cm}
\caption{\label{fig:selfenergy} Scalar self-energy correction to 
the forward-scattering amplitude.}
\end{figure}

It is instructive to discuss how the self-energy correction to the 
intermediate scalar in the full theory, see Figure~\ref{fig:selfenergy}, 
is represented in the effective 
description. We first separate the hard and soft contributions the 
one-loop self-energy, $\Pi(s)=\Pi_h(s)+\Pi_s(s)$, and then expand the 
hard part $\Pi_h(s) = \hat{M}^2 \sum_{l}\delta^l \, \Pi^{(1,l)}$. 
The soft part is reproduced by the effective theory self-energy. The 
first term $\Pi^{(1,0)}$ in the hard expansion is gauge-invariant 
and contributes to $\Delta^{(1)}$ as discussed before. This term is 
already relevant to the leading-order line-shape. The next term 
$\Pi^{(1,1)}$ cancels one of the adjacent scalar propagators, such that 
the self-energy correction merges with the local production vertex. 
$\Pi^{(1,1)}$ is gauge-dependent. The gauge dependence cancels with 
the vertex diagram to produce a gauge-independent NLO hard-matching 
coefficient $C^{(1)}$. Continuing in this way, we find that 
$\Pi^{(1,2)}$ contributes to the one-loop matching coefficient $D^{(1)}$ 
of the four-fermion operator $\left(\bar{\psi}_{n_-} \chi_{n_+}\right)  \!
  \left(\bar{\chi}_{n_+} \psi_{n_-}\right)$, because the scalar propagators 
to the left and right are both cancelled. The contribution is again required 
to obtained a gauge-invariant one-loop matching 
coefficient \cite{Beneke:2004km}, though it is already a NNLO term for 
the line-shape.

This example illustrates the power of the effective field theory method. 
It automatically breaks a diagram into different pieces and organizes 
them into gauge-invariant objects. The power-counting associated with 
the Lagrangian allows one to identify the terms relevant for a specified 
accuracy before any explicit calculation needs to be performed.

\subsection{Line-shape at next-to-leading order}

Our goal is to carry out this programme for the forward-scattering 
amplitude  ${\cal T}^{(0)} +{\cal T}^{(1)}$ at NLO, where 
${\cal T}^{(0)}$ sums up all terms that scale as $(g^2/\delta)^n\sim 1$ 
and ${\cal T}^{(1)}$ contains all terms that
are suppressed by an additional power of $g^2$ or $\delta$.
At leading order there is only one diagram, involving two
three-point vertices and one resonant scalar propagator. We get
\begin{equation}
\label{eq:LOT0}
i {\cal T}^{(0)} =
\frac{- i\, y y^*}{2\hat M {\cal D}} \,
[\bar{u}(p)v(q)]\,[\bar{v}(q)u(p)] ,
\end{equation}
where we defined ${\cal D}\equiv \sqrt{s}-\hat M-\Delta^{(1)}/2$. 
The inclusive line-shape is related to ${\cal T}^{(0)}$ by 
$\sigma = \mbox{Im}\,{\cal T}^{(0)}/s$ through the optical theorem. 
The above expression gives a Breit-Wigner distribution in 
$\sqrt{s}$.

\begin{figure}
\vskip0.2cm
\begin{center}
\includegraphics[width=7.5cm]{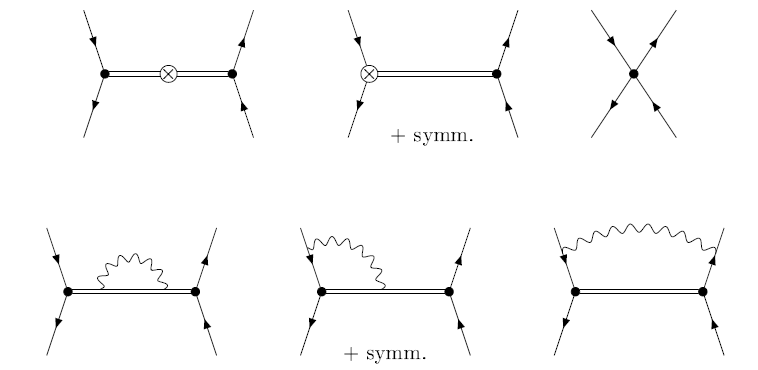}
\end{center}
\vspace*{-0.3cm}
\caption{\label{fig:NLO} Hard (upper) and soft (lower diagrams) 
contributions to ${\cal T}^{(1)}$.}
\end{figure}

In the effective theory there are three classes of diagrams that
contribute to ${\cal T}^{(1)}$, corresponding to hard, hard-collinear 
and soft contributions. The hard-collinear corrections to the external 
lines lead to scaleless integrals and vanish. The hard corrections
consist of a propagator insertion $[\Delta^{(1)}]^2/4 - \hat{M}
\Delta^{(2)}$, a production vertex insertion $C^{(1)}$, and a four-point 
vertex diagram due to the $(\bar{\psi}\chi)(\bar{\chi}\psi)$ 
operator in ${\cal L}_{\rm int}$, as shown in the upper diagrams of 
Figure~\ref{fig:NLO}. The sum of these diagrams reads
\begin{equation}
i\,{\cal T}^{(1)}_{h} = i\,{\cal T}^{(0)} \times
\Bigg[ 2 C^{(1)} - \frac{[\Delta^{(1)}]^2}{8{\cal D}\hat{M}} 
        + \frac{\Delta^{(2)}}{2{\cal D}}-\frac{{\cal D}}{2\hat{M}} 
\ \Bigg].\quad
\end{equation}
The soft-photon one-loop corrections (lower set of diagrams in 
Figure~\ref{fig:NLO}) computed in the effective theory result in 
\begin{equation}
i\,{\cal T}^{(1)}_{s} = i\,{\cal T}^{(0)} \times  \,\frac{g^2}{(4\pi)^2}\, 
\Bigg[4 L^2 - 4 L + \frac{5\pi^2}{6} \Bigg]
\quad
\label{eq:softNLO}
\end{equation}
with $L = \ln\left(-2{\cal D}/\mu\right)$. The partonic line-shape is 
obtained after subtracting the initial-state collinear
singularity and taking the imaginary part. 
The partonic line-shape must then be convoluted with the 
electron distribution function. 

We note the simplicity of the result, which is a consequence of the 
fact that the complete calculation is broken into separate single-scale 
calculations by factorizing the hard and soft regions. The NLO 
correction leads to a distortion of the line-shape relative to the 
Breit-Wigner form, which in non-inclusive situations can depend on the 
final state. Fitting a measured line-shape to the Breit-Wigner form 
rather than the true shape predicted by theoretical calculations 
leads to errors in mass determinations. In the present toy model, 
choosing the pole mass $M=100$~GeV (such that the $\overline{\rm MS}$ 
mass is $\hat{M}=98.8$~GeV at LO and $\hat{M}=99.1$~GeV at NLO) and 
couplings $g^2/(4\pi) = |y|^2/(4\pi) = 0.1$ to mimick the parameters 
of electroweak gauge bosons, the error would be of 
order 100 MeV.

\begin{figure}
\vskip0.2cm
\begin{center}
\includegraphics[width=7cm]{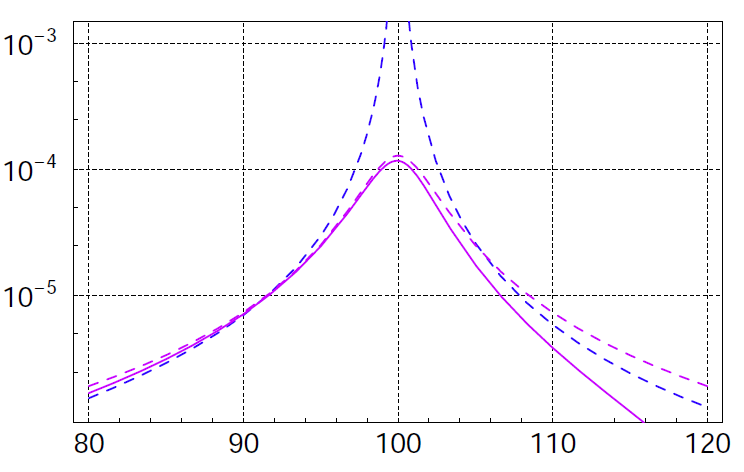}
\end{center}
\vspace*{-0.1cm}
\caption{\label{fig:ls} The line-shape (in GeV$^{-2})$ in the
effective theory at LO (light grey/magenta dashed) and NLO (light
grey/magenta) and the LO cross section off resonance in the full
theory (dark grey/blue dashed) as a function of the centre-of-mass
energy (in GeV). Figure from \cite{Beneke:2003xh}.}
\end{figure}

Figure~\ref{fig:ls} shows the leading-order partonic line-shape in the
effective theory and the tree-level (order $\alpha^2$) cross section
off resonance in the full theory. The two results agree in an
intermediate region where both calculations are valid. This allows to
obtain a consistent LO result for all values of $\sqrt{s}$. The figure also
shows the NLO line-shape for the numerical values given above. In order 
to obtain an improved NLO result in the entire 
region of $\sqrt{s}$, the NLO line-shape would have to be matched to
the NLO off-resonance cross section in the full theory.

The method discussed here makes NNLO  line-shape
calculations in $2\to2$ scattering possible with present techniques. 
An outline of such a calculation has been given in 
\cite{Beneke:2004km}, though no complete calculation has been 
performed to date.

\section{Pair production near threshold}

I reviewed in some detail the case of the line-shape, since it serves well 
to illustrate the general framework of unstable-particle effective 
theory. However, some of the more interesting 
results concern pair production of unstable particles, specifically the 
$W$ bosons and top quarks, near threshold. In $e^+ e^-$ collisions very 
precise measurement of the masses of these particles can be obtained from 
a threshold scan.

The threshold dynamics is determined by the interplay of the strength of 
the electromagnetic ($W$ bosons) or colour (top quarks) Coulomb force and 
the size of the decay width of the particle. The small parameters are 
\begin{equation}
\delta\equiv \frac{\Gamma}{M},\quad 
v^2 \equiv (\sqrt{s}-[2 M+i \Gamma])/M,\quad
\end{equation}
and the coupling $\alpha=g^2/(4\pi)$. For $W$ bosons, $\Gamma_W \sim M_W 
\alpha_{\rm EW}$ and therefore the effective strength of the Coulomb force 
is $\alpha_{\rm em}/v \sim \sqrt{\delta}\ll 1$. This leads to an 
enhancement, but the Coulomb force is never ${\cal O}(1)$, and no 
resummation is needed~\cite{Fadin:1995fp}. The rapid decay of the 
$W$ boson prevents the formation of any visible $W^+ W^-$ resonance. 
The situation is different for top quarks, since the Coulomb force 
is generated by QCD, while the decay still occurs 
through the electroweak interaction. Counting $\alpha_s \sim 
\alpha_{\rm EW}^2$, we now find  $\alpha_s/v \sim 1$. Diagrammatically, 
ladder diagrams that contain these enhanced terms must be summed to 
all orders in perturbation theory, which generates toponium bound-states 
in the spectral functions. Since the characteristic energy near threshold 
$E\sim M v^2$ is of order $\Gamma$, the bound-states appear as broad 
resonances, of which only the first one leaves a distinctive feature 
in the $t\bar t$ cross section~\cite{Fadin:1987wz,Fadin:1988fn}.

In the following I review results for $W$ and top pair 
production near threshold obtained within the effective field theory 
approach, leaving out all the technical details that 
can be found in the original papers.

\subsection{$W$ bosons}

This subsection summarizes results from \cite{Beneke:2007zg, Actis:2008rb}. 
We consider the process $e^- e^+ \to \mu^-\bar\nu_\mu u\bar{d} \,X$ 
with centre-of-mass energy $\sqrt{s} = 160 ... 170\,$GeV, where it is 
dominated by a $W^+ W^-$ intermediate state near threshold with 
subsequent semi-hadronic decay. The inclusive cross section is extracted from 
specific cuts of the forward amplitude
\begin{equation}
\hat \sigma =\frac{1}{s} \,\mbox{Im}\,{\cal A}(e^-e^+\to e^- e^+)_{| 
\mu^-\bar\nu_\mu u\bar{d}},
\end{equation}
which also includes diagrams with only a single internal $W$ line.
We perform a ``QCD-style'' calculation of the ``partonic'' 
cross section $\hat \sigma$ with massless electrons in the 
$\overline{\rm MS}$ scheme, and convolute it with the 
$\overline{\rm MS}$ electron distribution function:
\begin{equation}
\sigma(s) = \int_0^1 dx_1 dx_2 \,f_{e/e}(x_1) \,f_{e/e}(x_2)\, 
\hat\sigma(x_1 x_2 s).   
\end{equation}
The $\overline{\rm MS}$ electron distribution function depends on $m_e$, 
but not on $\sqrt{s}$, $M$, $\Gamma$. 

In the effective field theory (EFT) the $W$ bosons are described by 
two non-relativistic three-vector fields $\Omega^i_a$, where 
$a=\pm$ refers to the charge of the $W$. The HSET Lagrangian relevant 
to a single (scalar) unstable particle is replaced by the PNRQED
Lagrangian \cite{Pineda:1998kn}, generalized to the case of an unstable 
vector particle. The relevant terms are
\begin{widetext}
\begin{equation}
\hspace*{-0.65cm}
{\cal L}_{\rm PNRQED} = \sum_{a=\mp} \left[\Omega_a^{\dagger i} \left(
i D_s^0 + \frac{\vec{\partial}^2}{2 {M}_W} - \frac{\Delta}{2} \right)
\Omega_a^i
+  \Omega_a^{\dagger i}\,
\frac{(\vec{\partial }^2-{M}_W \Delta)^2}{8 {M}_W^3}\,
\Omega_a^i\right] +\!\int \!d^3 \vec{r}\, 
\left[\Omega_-^{\dagger i} \Omega^i_-\right]\!(x+\vec r\,)
\left(-\frac{\alpha_{\rm em}}{r}\right)
\left [\Omega_+^{\dagger j}\Omega^j_+\right]\!(x).\,
\label{LPNR}
\end{equation} 
\end{widetext}
\hspace*{-0.06cm} 
The master formula for the forward amplitude ${\cal A}$ coincides 
with (\ref{eq:master}), but the production and non-resonant operators 
are now of the form 
\begin{equation}
{\cal O}_p^{(k)} = C^{(k)}_{p}
\left(\bar{e}_{c_2,L/R} \gamma^{[i} n^{j]} e_{c_1,L/R} \right)
\left(\Omega_-^{\dagger i} \Omega_+^{\dagger j}\right) ,
\label{LPlead}
\end{equation}
\begin{equation}
\label{eq:4eLag}
{\cal O}^{(k)}_{\rm nr}= C^{(k)}_{\rm nr}\,
(\bar e_{c_1}\Gamma_1 e_{c_2})(\bar e_{c_2}\Gamma_2 e_{c_1}),
\end{equation}
with $\Gamma_1$, $\Gamma_2$ Dirac matrices, 
$a^{[i} b^{j]}\equiv a^i b^j + a^j b^i$, and $\vec{n}$ the unit-vector
in the direction of the incoming electron three-momentum.

Due to the $1/v$ enhancement of electromagnetic Coulomb exchange, the 
systematic expansion of ${\cal A}$ goes in powers of $\sqrt{\delta}$. 
Also, the non-resonant term appears as such a ``N$^{1/2}$LO'' 
correction, since the leading imaginary parts of $C^{(k)}_{\rm nr}$ are 
proportional to $\alpha^3$, while 
${\cal A} \sim \alpha^2\sqrt{\delta}$.\footnote{The factor $\sqrt{\delta}$ 
arises from the leading-order EFT matrix element and 
corresponds to the phase-space suppression near threshold.}

\begin{figure}
\vskip0.2cm
\begin{center}
\includegraphics[width=7cm]{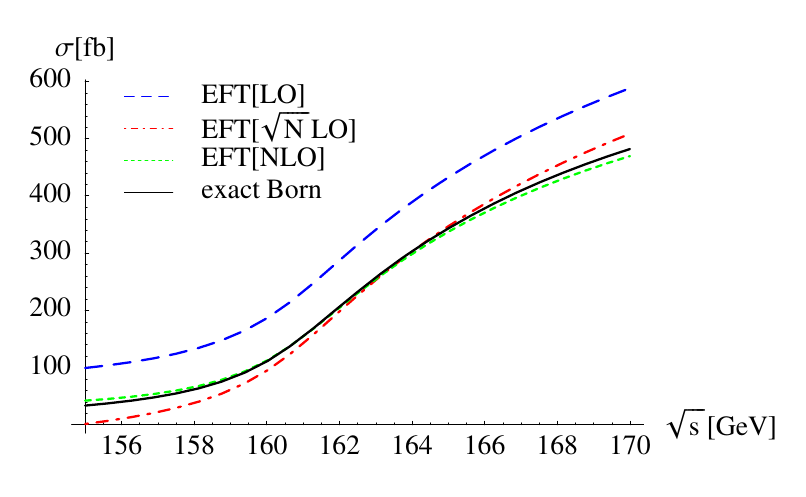}
\end{center}
\vspace*{-0.3cm}
\caption{Successive EFT approximations:
LO (long-dashed/blue), $\mbox{N}^{1/2}$LO (dash-dotted/red) and 
NLO (short-dashed/green). The solid/black 
curve is the full Born result computed with Whizard/ Comp\-Hep.
The $\mbox{N}^{3/2}$LO EFT approximation is indistinguishable from
the full Born result on the scale of this plot. 
Figure from \cite{Beneke:2007zg}.}
\label{fig:eft}
\end{figure}

The EFT constructs an expansion in $\Gamma/M$ and $(\sqrt{s}- 2M)/M$ 
of the full theory Born cross section. Before turning to radiative 
corrections it is instructive to compare successive terms in this 
expansion to the full Born result computed numerically (using 
Whizard \cite{Kilian:2007gr} and CompHep \cite{Pukhov:1999gg}). 
This is shown in Figure~\ref{fig:eft}. The LO non-relativistic approximation 
overestimates the true result. The N$^{1/2}$LO non-resonant correction 
yields a (nearly) constant, negative term and provides already 
good agreement close to the nominal threshold at $\sqrt{s}\approx 161\,$
GeV. To extend the approximation in a wider region around the threshold, 
it is necessary to include all terms up to N$^{3/2}$LO.

\begin{figure}
\vskip0.2cm
\begin{center}
\includegraphics[width=2.5cm]{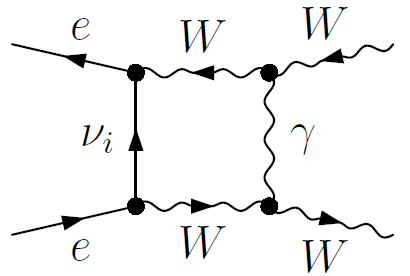}
\hskip0.5cm
\includegraphics[width=2.5cm]{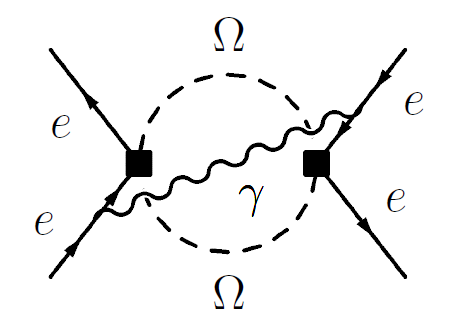}
\end{center}
\vspace*{-0.3cm}
\caption{One-loop diagram for the hard-matching coefficient (left) 
and a soft NLO contribution to the forward-scattering amplitude in 
the effective theory (right).}
\label{fig:wwdiags}
\end{figure}

While constructing an expansion of the Born cross section when an 
exact, numerical result is readily available, appears as an unnecessary 
complication, the computation of the NLO radiative correction in 
unstable-particle effective theory \cite{Beneke:2007zg} is 
remarkable simple compared to the corresponding calculation in the 
complex-mass scheme \cite{Denner:2005fg,Denner:2005es}. The most 
complicated part is the computation of the NLO matching 
coefficient of the operator (\ref{LPlead}), which, however, is a standard 
one-loop calculation. A representative diagram is shown in 
Figure~\ref{fig:wwdiags} left. The diagram on the right displays 
a soft, ``non-factorizable'' NLO correction to the two-point 
function of production operators in (\ref{eq:master}), and results 
again in a simple expression, similar to (\ref{eq:softNLO}). 
A comparison with the complex-mass scheme calculation and the 
double-pole approximation (DPA), including 
QCD corrections and initial-state radiation is given in the 
following table.\footnote{The ``full ee4f'' column refers to the 
erratum of \cite{Denner:2005es}.} The numerical difference of 
1\% between the EFT and full ee4f results is presumably in part due to 
the N$^{3/2}$LO correction associated with the NLO matching coefficient 
of ${\cal O}^{(k)}_{\rm nr}$, which is implicitly contained in the 
NLO full ee4f calculation.

\begin{center}
{\footnotesize \begin{tabular}{|c|c|c|c|c|}
\hline&
\multicolumn{3}{c}{
$\sigma(e^-e^+\to \mu^-\bar\nu_\mu u\bar d\,X)$(fb)}&
\\\hline
$\sqrt{s}\,[\mbox{GeV}]$& Born (SM) & EFT & full ee4f & DPA \\\hline
161 &  107.06(4) &117.38(4)   & 118.77(8)& 115.48(7)\\\hline
170 &  381.0(2) & 399.9(2)   & 404.5(2) & 401.8(2)  \\\hline
\end{tabular}
}
\end{center}

\begin{figure}
\vskip-0.2cm
\begin{center}
\includegraphics[width=7cm]{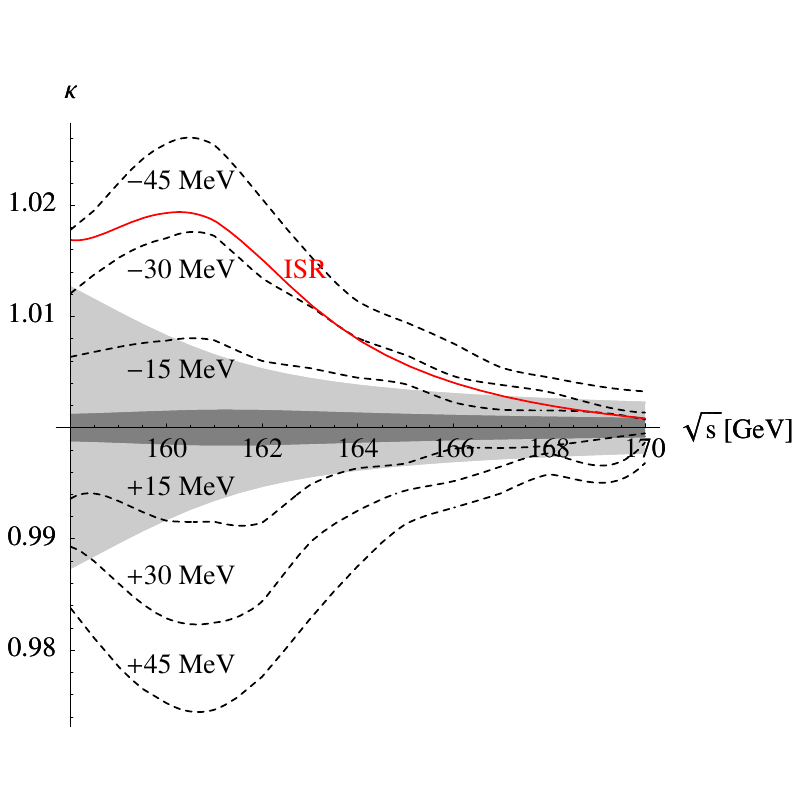}
\end{center}
\vspace*{-0.5cm}
\caption{$W$-mass dependence of the total cross section. All the
cross sections are normalized to $\sigma(s,M_W=80.377\,\mbox{GeV})$.
See text for explanations. Figure from \cite{Beneke:2007zg}.}
\label{fig:massuncertainty}
\end{figure}

We can now estimate the theoretical uncertainty in the $W$ mass 
determination from a threshold scan. Figure~\ref{fig:massuncertainty} shows 
$\kappa = \sigma(s,M_W+\delta M_W)/\sigma(s,M_W)$ for $M_W=80.377\,\mbox{GeV}$ 
and different values of $\delta M_W$ as function of the cms energy. 
The relative change in
the cross section is shown as dashed lines for $\delta M_W=\pm
15,\pm 30, \pm 45 \, \mbox{MeV}$. The shape of these curves shows
that the sensitivity of the cross section to the $W$ mass is
largest around the nominal threshold $\sqrt{s}\approx 161\,$GeV,
as expected, and rapidly decreases for larger $\sqrt{s}$. (The loss
in sensitivity is partially compensated by a larger cross section,
implying smaller statistical errors of the anticipated experimental
data.) The shaded areas provide an estimate of the uncertainty 
from uncalculated N$^{3/2}$LO terms. The inner band is associated with the 
interference of single Coulomb exchange with one-loop hard or soft 
corrections, which are genuine NNLO corrections in other schemes. 
The outer band accounts for the non-resonant term already mentioned 
above. Finally, the line marked ``ISR'' estimates ambiguities 
in the implementation of initial-state radiation. This represents 
the largest current uncertainty. In order to obtain 
a competitive determination of $M_W$, one eventually needs a more 
accurate computation of the electron distribution function. 

\begin{figure}
\vskip0.2cm
\begin{center}
\includegraphics[width=2.3cm]{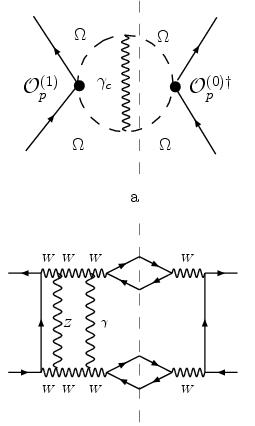}
\includegraphics[width=4.6cm]{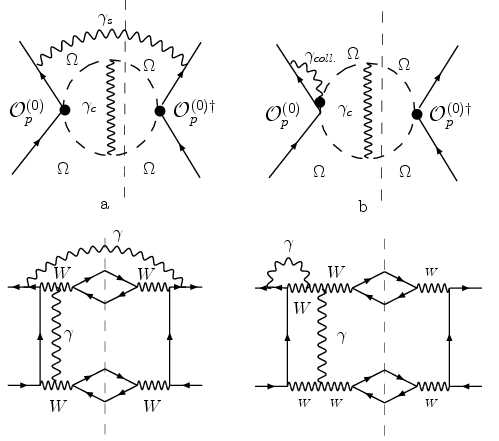}
\end{center}
\vspace*{-0.5cm}
\caption{Some NNLO diagrams that count as N$^{3/2}$LO 
in the $\delta$ expansion and their EFT representation 
(upper line).}
\label{fig:nnlo}
\end{figure}

Since this is not a fundamental problem and since the full theory NLO ee4f 
calculation is available, the accuracy of the theoretical prediction 
is limited by the N$^{3/2}$LO terms in the $\delta$ expansion, which 
correspond to two-loop corrections (in the complex-mass scheme). Some of 
the diagrams together with their EFT representation are shown in 
Figure~\ref{fig:nnlo}. These consist of mixed hard-Coulomb corrections 
(first column), interference of Coulomb exchange with soft and collinear 
radiative corrections (2nd and 3rd column, respectively), and 
a correction to the electromagnetic Coulomb potential itself. These 
genuine higher-order corrections have been computed 
\cite{Actis:2008rb} and were found to be below 0.5\%, leading to 
shifts of $W$ mass of less than 5 MeV.

\begin{figure}
\vskip0.2cm
\begin{center}
\includegraphics[width=7cm]{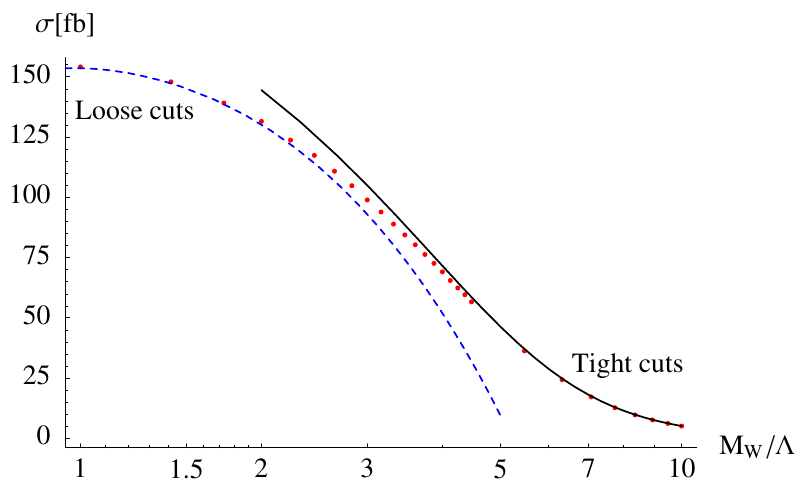}
\end{center}
\vspace*{-0.5cm}
\caption{Comparison of the Born cross section in the full SM at $\sqrt s =161$ GeV computed 
with WHIZARD (red dots) with the effective-theory result for the 
loose-cut implementation (dashed blue curve) and the tight-cut implementation
(solid black curve). Figure from \cite{Actis:2008rb}.}
\label{fig:nonresW}
\end{figure}

Up to now, we considered the total cross section for the flavour-specific 
final state $\mu^-\bar\nu_\mu u\bar{d} \,X$. Since experimentally 
certain cuts must be applied, it would be desirable to compute directly 
the cut cross section in the EFT. A framework to implement arbitrary 
cuts while maintaining an expansion in the power-counting parameter 
$\delta$ is not available and probably difficult to achieve. The 
specific case of invariant-mass cuts $|M_{f_if_j}^2-M_W^2| <
\Lambda^2$ on the $W$-decay products has been considered in 
\cite{Actis:2008rb}. The implementation depends on how $\Lambda$ 
scales with the parameter $\delta$. For loose cuts, $\Lambda \sim 
M_W$. Since by assumption the virtualities in the EFT are at most of 
order $M \Gamma \sim M\delta$, the loose cut does not affect the EFT 
diagrams. However, the hard-matching coefficients are modified and 
acquire a dependence on $\Lambda$ in addition to the other short-distance 
scales. The situation is reversed for tight cuts with 
$\Lambda \sim M \Gamma \sim M\sqrt{\delta}$. The tight cut cuts into 
the (approximate) Breit-Wigner distribution of the single-$W$ 
invariant mass distribution and therefore must be applied to the 
calculation of the EFT loop integrals. On the other hand, it eliminates 
off-shell contributions, and hence the short-distance coefficient 
$C^{(k)}_{\rm nr}$ of the non-resonant four-electron operator 
(\ref{eq:4eLag}) vanishes. Figure~\ref{fig:nonresW} shows good agreement 
of the effective-theory calculation of the cut Born cross section with the 
numerical result from WHIZARD in the regions where the respective
loose/tight-cut counting rule is appropriate.

\subsection{Top quarks}

\begin{figure}
\vskip0.2cm
\begin{center}
\includegraphics[width=7cm]{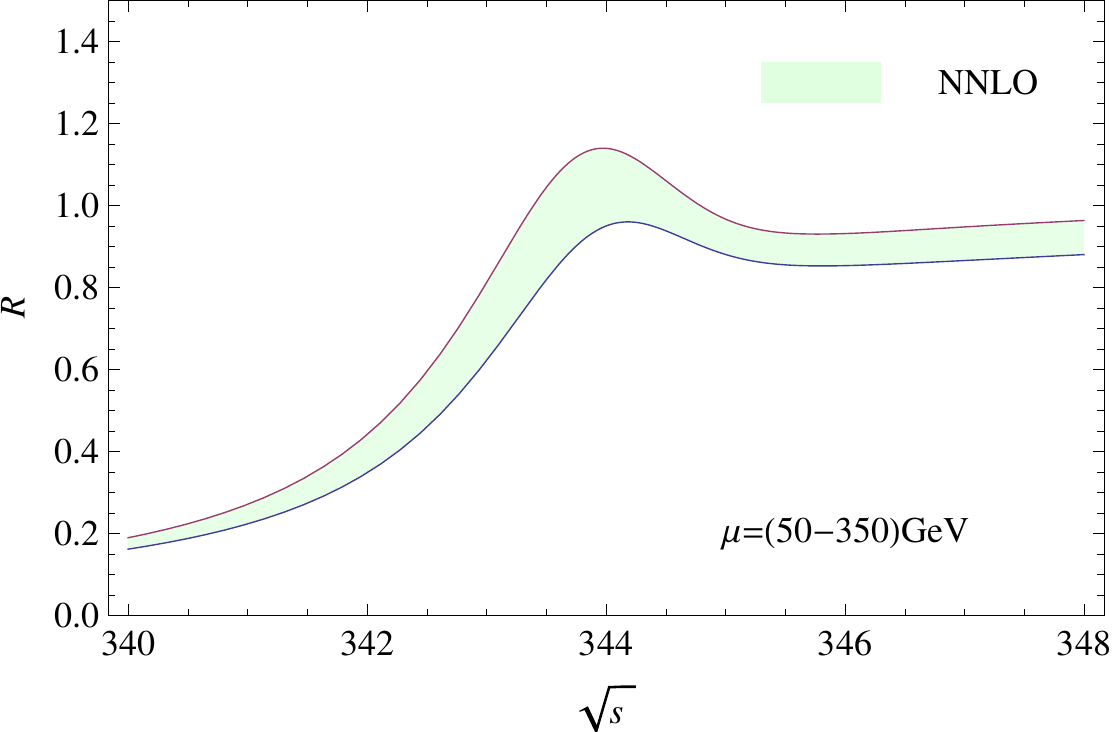}
\end{center}
\vspace*{-0.3cm}
\caption{Top quark pair production cross section near threshold, 
normalized to $4\pi\alpha_{\rm em}^2/(3 s)$ at NNLO in  
non-relativistic, resummed perturbation theory in the PS mass 
scheme \cite{Beneke:1998rk}. The width of the band reflects the 
theoretical uncertainty estimated from scale variation in the 
indicated range. 
$m_{t,\rm PS}(20\,\mbox{GeV})=171.5\,$GeV, $\Gamma_t=1.33\,$GeV.}
\label{fig:topNNLO}
\end{figure}

The strong Coulomb attraction of the coloured top quarks, when their 
relative velocity is small near threshold, requires the  
resummation of certain QCD corrections to all orders in perturbation 
theory. A systematic formalism employs a sequence of matching 
steps to define hard matching coefficients and potentials in 
non-relativistic effective field theory. The ingredients required 
up to the third order in non-relativistic perturbation theory are 
reviewed in \cite{Beneke:2013jia}. Figure~\ref{fig:topNNLO} shows 
the second-order (NNLO) result from~\cite{Beneke:1999qg}, which exhibits 
a toponium resonance slightly below the nominal threshold. 
Note the order counting here is such that $v\sim \alpha_s$ defines 
the expansion parameter. The top-quark width $\Gamma_t/m_t\sim 
\alpha_{\rm EW} \sim \alpha_s^2$ is second-order in this counting.

QCD predictions of the top-pair production cross section such as the 
one shown in the figure are based on 
the calculation of QCD correlation functions with the substitution 
$E=\sqrt{s}-2 m_t \to E+i\Gamma_t$ to account for the top-quark 
width~\cite{Fadin:1987wz,Fadin:1988fn}. This prescription corresponds 
to computing the first (resonant) term in (\ref{eq:master}) with 
${\cal O}_p^{(k)}$ given by the non-relativistic top-quark (axial-) vector 
current, and with  
an effective Lagrangian that accounts for the width through 
$\Delta  = - i\Gamma_t$ in the first bilinear term in (\ref{LPNR}) 
(adapted to quarks), but not in the further kinetic corrections.

The limitations of this approximation manifest themselves within 
the (NR)QCD calculation itself. The current correlation function $G(E)$ 
exhibits an uncancelled ultraviolet divergence from an overall 
divergence of the form $[\delta G(E)]_{\rm overall} 
\propto \alpha_s E/\epsilon$ in dimensional regularization 
($d=4-2\epsilon$) \cite{Beneke:2008cr}. Since $E$ acquires an 
imaginary part $\Gamma_t \sim m_t\alpha_{\rm EW}$, the divergence 
survives in the cross section,
\begin{equation} 
\mbox{Im}\,[\delta G(E)]_{\rm overall} \propto 
m_t\times \frac{\alpha_s\alpha_{\rm EW}}{\epsilon},
\end{equation}
and appears first at NNLO (since at LO, $G(E) \sim v \sim \alpha_s$).
A consistent calculation therefore requires that one considers 
the process $e^+ e^- \to W^+W^- b\bar b$ within unstable-particle 
effective theory including the effects of off-shell top quarks and 
processes that produce the $W^+W^- b\bar b$ final state with no or only one 
intermediate top-quark line. The two terms of (\ref{eq:master}) can be 
identified with 
\begin{widetext}
\begin{equation}
\hskip 3cm \sigma_{e^+ e^- \to W^+W^- b\bar b} = 
\underbrace{\sigma_{e^+ e^- \to [t\bar t]_{\rm res}}(\mu_w)}_{
\mbox{\footnotesize pure (NR)QCD}} + \,
\sigma_{e^+ e^- \to W^+W^- b\bar b{}_{\rm nonres}}(\mu_w). 
\end{equation}
\end{widetext}
\hspace*{0.04cm} 
Both terms separately have a ``finite-width scale dependence'' related 
to the uncancelled $1/\epsilon$ poles, and only the sum is well-defined. 
For consistency, both terms have to be defined with the same 
(dimensional) regularization prescription. 

While the explicit finite-width scale dependence is seen first at 
NNLO, the leading non-resonant contribution already appears at NLO. 
Somewhat surprisingly, this was realized only recently. At this order 
the matching coefficient of the non-resonant operator is 
equivalent to the dimensionally regulated 
$e^+ e^- \to b W^+\bar t$ process with $\Gamma_t=0$, 
expanded in the hard region around $s=4 m_t^2$. The corresponding 
calculation has been performed in two independent 
ways~\cite{Beneke:2010mp,Penin:2011gg}. In \cite{Beneke:2010mp} 
invariant-mass cuts $m_t -\Delta M_t   \le M_{t,\bar{t}} 
\le m_t +\Delta M_t$
on the decay products of the (anti-) top quark can be included 
following the method discussed above for $W$ bosons. 

\begin{figure}
\vskip0.2cm
\begin{center}
\includegraphics[width=4.5cm]{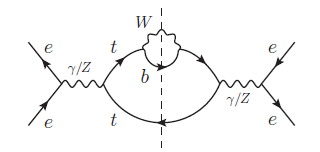}
\end{center}
\vspace*{-0.3cm}
\caption{Hard, off-shell top diagram contribution to 
$e^+ e^- \to b W^+\bar t$, which leads to a linear infrared divergence.}
\label{fig:diagh1}
\end{figure}

It is convenient 
to represent the calculation of the cut two-loop diagrams 
contributing to $e^+ e^- \to b W^+\bar t$ in the form 
\begin{equation}
\int_{\Delta^2}^{m_t^2} dp_t^2 \,(m_t^2-p_t^2)^{\frac{d-3}{2}} 
H_i\Big(\frac{p_t^2}{m_t^2},\frac{M_W^2}{m_t^2} \Big)
\label{eq:ptint}
\end{equation}
leaving the integration over $p_t^2 \equiv (p_b+p_{W^+})^2$ to the 
end. The lower limit depends on the invariant-mass cut and is given 
by $\Delta = M_W^2$, when no cut is applied.
The finite-width infrared divergence of the non-resonant matching 
coefficient that cancels the corresponding ultraviolet divergence 
of the non-relativistic current correlation function appears as 
an endpoint divergence of the integral above as $p_t^2 \to 
m_t^2$ approaches the on-shell value. At NLO, the divergence arises 
only from the diagram shown in Figure~\ref{fig:diagh1}. The 
integrand behaves as 
\begin{equation}
H_1\Big(\frac{p_t^2}{m_t^2},\frac{M_W^2}{m_t^2} \Big) 
\;\;\stackrel{p_t^2\to m_t^2}{\rightarrow} \;\;\mbox{const} \times 
\frac{1}{(m_t^2-p_t^2)^2},
\end{equation}
which leads to a divergent integral (\ref{eq:ptint}) in four dimensions. 
Dimensional regularization sets such linearly divergent integrals to 
finite numbers, which explains the absence of explicit $\mu_w$ scale 
dependence at this order. 

\begin{figure}
\vskip0.2cm
\begin{center}
\includegraphics[width=7cm]{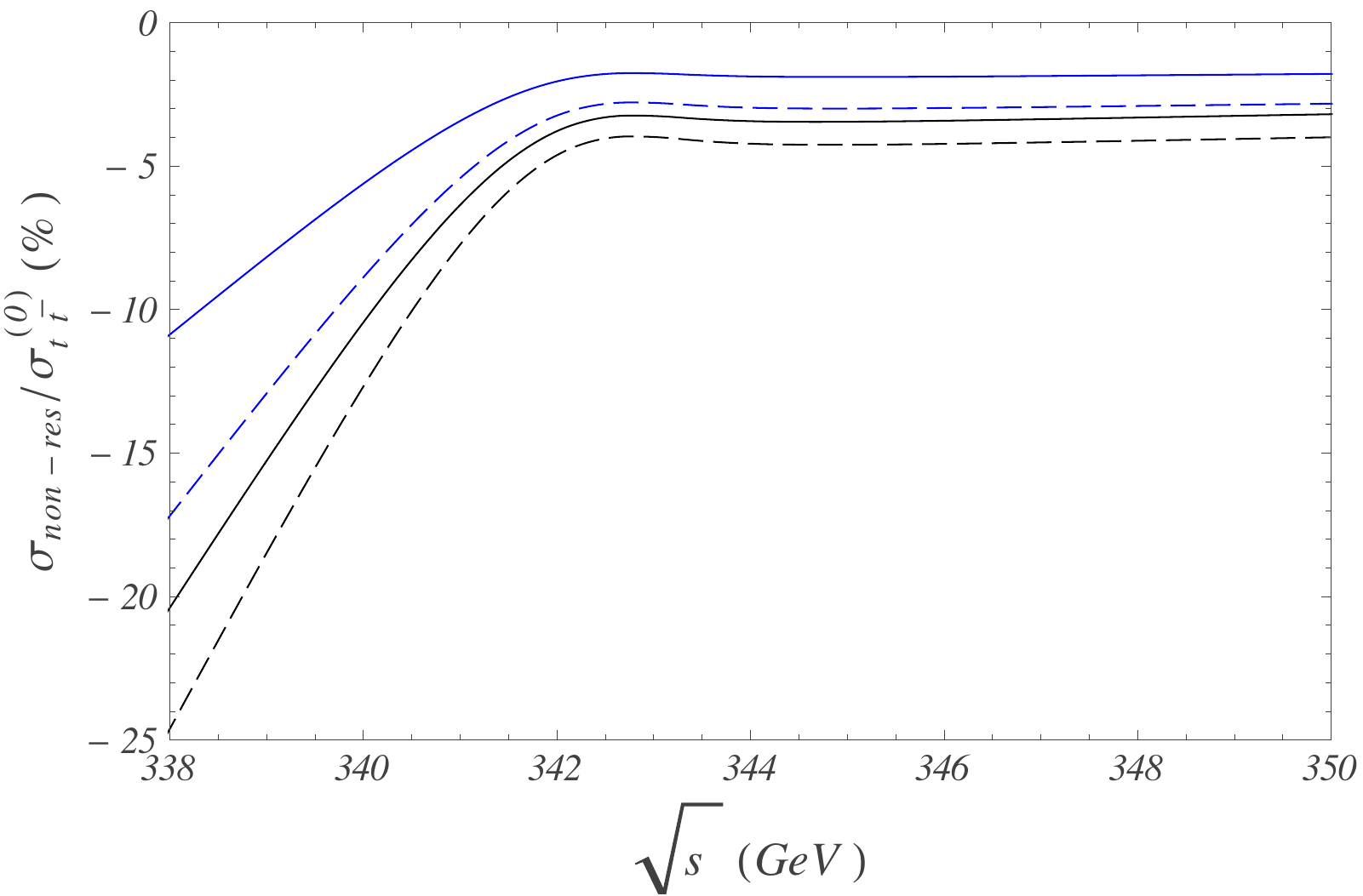}
\end{center}
\vspace*{-0.3cm}
\caption{Relative sizes of the non-resonant corrections with respect to
the $t\bar{t}$ LO cross section in percent:  NNLO singular 
terms $\sigma_{\rm non-res}^{(2)}/\sigma_{t\bar{t}}^{(0)}$
(upper blue lines) and NLO 
$\sigma_{\rm non-res}^{(1)}/\sigma_{t\bar{t}}^{(0)}$ 
(lower black lines). Solid (dashed) lines 
correspond to an invariant-mass cut $\Delta M_t=35$~GeV 
($\Delta M_t=15$~GeV). Figure from \cite{Jantzen:2013gpa}. 
Here the top pole mass $m_t=172\,$GeV is used as input parameter.}
\label{fig:topNNLOnonres}
\end{figure}

Similar to the case of $W$-boson pair production, the leading non-resonant 
contribution to the top-pair cross section is a nearly energy-independent, 
negative correction, which amounts to a few percent above the toponium 
peak, and to around 20\% a few GeV below the peak. With an invariant-mass 
cut the NLO correction is shown as the lower (black) solid and dashed lines 
in Figure~\ref{fig:topNNLOnonres}.

Since the largest sensitivity to the top-quark mass comes from 
the steep rise of the cross section below the peak, the non-resonant 
contributions at NLO and even NNLO are essential. The NNLO terms 
correspond to three-loop cut diagrams. The complete calculation has not 
yet been performed. However, in the presence of an invariant-mass cut 
satisfying $\Gamma_t \ll \Delta M_t \ll m_t$, the singular terms 
as $m_t/\Delta M_t \to \infty$ have been extracted in two different 
ways \cite{Jantzen:2013gpa,Hoang:2010gu}.\footnote{Further, the leading 
term in an expansion in the parameter $\rho=1-M_W/m_t$ has been obtained 
in \cite{Penin:2011gg} and \cite{Ruiz-Femenia:2014ava}, with different 
results.} The calculation of 
\cite{Jantzen:2013gpa}, which starts from the non-resonant side, 
also confirms explicitly the cancellation of $1/\epsilon$ finite-width 
divergence poles with the non-relativistic contributions. The upper set 
of lines (blue solid and dashed) in Figure~\ref{fig:topNNLOnonres}, 
shows that the NNLO correction is only about half as large than the 
NLO one.

\section{Summary and further results}

In this article I reviewed the treatment of unstable particles in 
perturbative quantum field theory based on the scale hierarchy 
$\Gamma \ll M$. Once scale separation is taken as the guiding principle, 
and an effective field theory is constructed, gauge-invariance and the 
consistency of the all-order resummation is automatic. The effective theory 
describes the scattering processes that leave the resonance close 
to its complex mass shell. Some aspects are therefore closely related 
to other effective theories that describe heavy particles close to their 
mass-shell. The concrete applications considered so far can be described 
by the master formula (\ref{eq:master}), which captures resonant 
production and decay, as well as all non-resonant ``background'' 
processes. 

The effective theory approach appears most fruitful, when it is applied 
to inclusive quantities, where it leads to particularly simple, even 
completely analytic results; to processes that require other resummations 
on top of the self-energy of the unstable particle; and to processes 
where high precision is required, beyond NLO accuracy, for which automated 
tools are not yet available. The line-shape and pair production near 
threshold discussed here are examples of such situations.

As with other effective theories it is more difficult to predict 
differential distributions, unless the scales associated with the 
observable can be assigned a unique scaling with respect to the small 
parameters that 
define the EFT expansion. To circumvent this problem, a hybrid approach 
has been followed in \cite{Falgari:2010sf,Falgari:2013gwa}, applicable 
to NLO calculations, in which the simplifications provided by 
the EFT are used in the virtual corrections (which have tree kinematics), 
while real emission is computed in the full theory with the complex-mass 
prescription. A fully differential calculation has been done recently 
for the mixed ${\cal O}(\alpha_{\rm em}\alpha_s)$ corrections to 
Drell-Yan production \cite{Dittmaier:2014qza} employing a mixture of 
diagrammatic and EFT-inspired techniques.

\subsection*{Acknowledgements}

\noindent
This review summarizes work performed within and supported by the 
DFG Sonderforschungsbereich/Trans\-regio~9 
``Computational Theoretical Particle Physics''. I wish to thank my 
collaborators on this project, in particular P.~Falgari, B.~Jantzen, 
P.~Ruiz-Femenia, Ch.~Schwinn, A. Signer and G.~Zanderighi.




\bibliographystyle{elsarticle-num}

\end{document}